\documentclass[nofootinbib, reprint, amsmath,amssymb, aps,]{revtex4-2}
\usepackage{graphicx}
\usepackage{dcolumn}
\usepackage{bm}
\usepackage[colorlinks]{hyperref}
\usepackage{soul}

\begin{document}

\title{Effective Field Theories in Magnetohydrodynamics}


\author{Amir Jafari}
\email{elenceq@jhu.edu}

\begin{abstract}
We briefly review the recent developments in magnetohydrodynamics, which in particular deal with the evolution of magnetic fields in turbulent plasmas. We especially emphasize (i) the necessity of renormalizing equations of motion in turbulence where velocity and magnetic fields become Hölder singular; (ii) the breakdown of Laplacian determinism (spontaneous stochasticity) for turbulent magnetic fields; and (iii) the possibility of eliminating the notion of magnetic field lines, using instead magnetic path lines as trajectories of Alfv\'enic wave-packets. These methodologies are then exemplified with their application to the problem of magnetic reconnection---rapid change in magnetic field pattern that accelerates plasma---a ubiquitous phenomenon in astrophysics and laboratory plasmas. The necessity of smoothing out rough velocity and magnetic fields on a finite scale $l$ implies that magnetohydrodynamic equations should be regarded as {\it{effective field theories}}
 with running parameters depending upon the scale $l$.
\end{abstract}
\pacs{Valid PACS appear here}
\maketitle

\section{Introduction}
Experiment is the ultimate guide in any physical theory. In assigning an electric charge $e$ to the electron, for instance, we must worry about the "measurability" of this charge in the laboratory. It turns out that what we measure for the fundamental charge $e$ indeed depends on the scale of measurement. Energy fluctuations  in vacuum, due to uncertainty principle, may be thought of as creation and annihilation of virtual particles because of mass-energy equivalence. Thus the electron is thought as surrounded by an infinite number of virtual particles, e.g., electrons and positrons, which "screen" the electron's charge making it a scale-dependent quantity---we measure different values for the electron's charge as we change the scale and get closer to the electron! The measured electric charge is the {\it{renormalized}} charge whereas what we initially assign as electric charge in equations of motion is the {\it{bare}} charge. This phenomenon is closely related to the theory of Renormalization Group (RG), which is widely used in quantum field theories and also classical theories. As an example for the latter, in classical theories of liquid diffusion \citep{donev2014reversible}, the bare diffusivity can even vanish while the renormalized diffusivity \cite{BedeauxMazur1974} is what satisfies the Stokes-Einstein relation and can be measured experimentally \citep{EyinkJafari2022}.\footnote{In fact, L. F. Richardson \cite{Richardson1926} already in 1926 found that the "effective diffusivity" $K:=\langle |\Delta {\bf x}|^2\rangle/t$ of soot particles in a turbulent volcanic plume scaled as $l^{4/3}$ with the distance $l=\langle |\Delta {\bf x}|^2\rangle^{1/2}$ between particles.} What can be measured and the error made in any measurement play even more crucial roles in chaotic and super-chaotic (turbulent) systems where the dynamics is extremely sensitive to initial conditions; see \S\ref{renormalization}. In fact, in turbulent flows, even the determinism of classical physics can be lost at scales much larger than any scale on which any quantum effect might have an effect, see e.g., \citep{Eyink2011, ELV2011, EyinkBandak2020}, as we will discuss in \S\ref{SSS}. To exemplify the importance and utility of these concepts, we will briefly discuss magnetic reconnection in \S\ref{sreconnection}, in a formalism independent of the notion of field lines \cite{Jafari2024}.

To appreciate the role of finite measurement scales in a magnetized plasma, we note that magnetic field measured at a point $\bf x$ is in fact an average field ${\bf B}_l$ in a ball of spatial scale $l>0$ around $\bf x$, not the "bare" field $\bf B$ defined at a dimensionless point which is only a mathematical idealization. Hence, equations of motion should be understood as renormalized or effective equations on finite measurement scales $l>0$. This simple fact becomes crucial in turbulence, where upon averaging, the equations of motion e.g., the induction equation, acquire non-linear terms which increase with decreasing the measurement scale in the inertial range, thus playing a dominant role in the dynamics. For instance, the non-linear turbulent term in the induction equation {\it{dominates}} over all small-scale plasma effects \cite{Eyink2015, JV2019}, rendering non-turbulent reconnection models totally unrealistic in astrophysics \citep{Lazarianetal2019, Review2020}. In turbulence, both magnetic and velocity fields become Hölder singular with ill-defined spatial derivatives \citep{Eyink2018, JV2019, Eyink2024}. Hölder singularity of the velocity (magnetic) field is intimately related to the Richardson diffusion \cite{Richardson1926} of Lagrangian particles (Alv\'enic wave-packets), which implies a super-chaotic behavior known as spontaneous stochasticity \cite{Bernardetal.1998} in which solutions of the equations of motion become non-unique \cite{Eyink2011, EyinkBandak2020} and initial conditions are completely forgotten! This is totally different from simple chaos (i.e., the butterfly effect) in which solutions diverge exponentially but initial conditions are never forgotten. A renormalization procedure can be applied to such turbulent fields \citep{EyinkAluie2006, Eyink2015, JV2019, EyinkBandak2020, Mizerski2021} to obtain well-defined equations of motion with unique solutions at large (inertial) scales; see also \cite{Canet2022} and references therein. This process involves averaging over (or integrating out) the small degrees of freedom below a given scale $l$.  
In such a description for systems with strong fluctuations at all length scales, encountered in quantum field theories and turbulent flows, the effective description of the system varies with the resolution length scale $l$. Instead of seeking for some idealized theory valid at vanishingly small length scales, in an "effective theory", one attempts to understand how the description changes as $l$ is varied \citep{Eyink2024}.

 Astrophysical flows usually have large Reynolds numbers, which indicates the presence of turbulence \cite{Burkhart2010} where magnetic field undergoes super-diffusion \citep{JVV2019, Review2020} and constantly reconnects everywhere on all scales \citep{LV99, ELV2011, Lazarianetal2019, Review2020}. As a main feature of magnetohydrodynamic (MHD) turbulence, this is inconsistent with the idea of magnetic field lines considered as persistent dynamical elements evolving smoothly with time \citep{JV2019, Review2020}. Trajectories of Alfv\'enic wave-packets, or magnetic path lines, provides an alternative tool to study the evolution and reconnection of magnetic fields. Magnetic topology change and its possible connection with reconnection can also be approached using path lines in a much simpler way \cite{Jafari2024}. Another important consequence of turbulence is the dramatic violation of flux-freezing in the presence of turbulence \citep{EyinkAluie2006, Eyink2011, Eyinketal2013, JVV2019}. 

Despite the fundamental importance of above mentioned concepts, they seem to remain unappreciated especially in plasma physics community. For example, both astrophysical observations as well as numerical simulations are in excellent agreement with turbulent reconnection theory, originally proposed by Lazarian and Vishniac \cite{LV99} and later refined mathematically in a series of works by Eyink and others; e.g., \citep{Eyink2015, Lazarianetal2015} based on the methodologies discussed above; see 
\citep{Kowal2009, Kowal2012, Eyinketal2013, Beresnyak2017, Oishietal2015, Kowal2017} for non-relativistic regime and e.g., \citep{TakamotoLazarian2016, Takamoto2018} for relativistic regime.
Nevertheless, plasma physics community has essentially neglected the whole theory and its implications for decades. Another overlooked phenomenon is the breakdown of magnetic flux freezing \cite{Alfven1942} in real astrophysical flows due to turbulence as mentioned above \citep{Lazarianetal2015, Lazarianetal2019, Review2020, Pontin2022}. One reason for these important developments to be neglected might be the unfamiliarity of the community with concepts such as effective field theories and renormalization group analysis, which are well-established concepts in other areas such as high energy and statistical physics. Thus to make this short review accessible to a larger audience, we will keep the technical level as simple as possible, referring the interested reader to the original literature for more detailed discussions. 

In the present review, we will first briefly discuss the concept of finite measurement scales and renormalization group analysis in \S\ref{renormalization}. The breakdown of Laplacian determinism in turbulence, i.e., spontaneous stochasticity, is discussed in \S\ref{SSS}. Finally, to exemplify the introduced concepts, we briefly revisit the problem of magnetic reconnection in \S\ref{sreconnection} in a formalism independent of magnetic field lines.

\section{Renormalization}\label{renormalization}
Any physical measurement can be performed only on a finite scale $l>0$ in space, which can be made smaller but it cannot vanish. Equations of motion, e.g., the Navier-Stokes and induction equations for velocity and magnetic fields, are in fact "effective" equations governing average, measurable quantities. Although with changing the scale, the value of physical quantities change, but they are all governed by the same physics, i.e., same equations of motion: decreasing the scale $l$ will make measurements "finer". Mathematically, to obtain an average, coarse-grained, or renormalized version of any field ${\bf F(x},t)$ on scale $l$, one may integrate it with a rapidly decaying kernel; 
\begin{equation}\label{coarsegrain1}
{\bf{F}}_l ({\bf{x}}, t)=\int_V \phi\left({{\bf{r}}\over l}\right).{\bf F}({\bf{x+r}}, t) {d^3r\over l^3},
\end{equation}
where $\phi({\bf{r}})=\phi(r)$ is a smooth and rapidly decaying (scalar) kernel (for mathematical properties required to be satisfied by this kernel, see e.g., \cite{Eyink1996}, Sec. 2.1).
Hence, ${\bf F}_l({\bf x},t)$ is what an experimentalist can measure for quantity ${\bf F(x},t)$ in a ball of size $l$ around point $\bf x$.\footnote{Spatial complexity of a vector field $\bf F$ can then be defined using the "angle" between ${\bf F}_l$ and ${\bf F}_L$ at measurement scales $l<L$. A spatially smooth (untangled) vector field would point almost in the same direction at any resolution unlike a tangled field. Thus spatial complexity (of order $p$ in a volume $V$ can be defined as $\Big[\int_V d^3x|1-\hat {\bf F}_l.\hat{\bf F}_L|^p\Big]^{1/p}$ where ${\bf F}:={\bf F}/|{\bf F}|$ is the unit direction vector \cite{JV2019}. This might be a useful statistical quantity in problems such as magnetic reconnection where magnetic field direction undergoes rapid changes \cite{Jafari2020}.}

Let us take the Ohm's law, ${\bf E+v\times B}-\eta {\bf J}=0$, as an example, with electric, magnetic and velocity fields $\bf E, B, v$, resistivity $\eta$ and current $\bf J=\nabla\times B$. The implication is that the physically measurable electric fields and electric current on any given measurement scale $l>0$, denoted by subscript $l$, are governed by ${\bf E}_l +  ({\bf v\times B})_l -\eta {\bf J}_l=\bf 0$. We can of course measure the velocity ${\bf v}_l$ as well as magnetic field ${\bf B}_l$ on this scale, and write the Ohm's law in its original form as
\begin{eqnarray}\label{Ohm2}{\bf E}_l + {\bf v}_l\times   {\bf B}_l-\eta   {\bf J}_l&=&
  {\bf v}_l\times   {\bf B}_l-  ({\bf v \times B})_l\\\nonumber
  &:=&{\bf R}_l,
  \end{eqnarray}
in terms of the turbulent electric field induced by motions at scales $<l$, represented by the non-linear term ${\bf R}_l$, which scales as $|{\bf R}_l|\sim |{\delta \bf v}(l) \times {\delta \bf B}(l)|$ in terms of velocity and magnetic field increments (measured across distance $l$). As long as the velocity and magnetic fields are continuous functions with well-defined spatial derivatives, i.e, they are Lipschitz continuous, $|{\delta \bf v}(l)| \propto l$ and $|{\delta \bf B}(l)| \propto l$.\footnote{For a single-variable function $f(x)$, this can be easily seen by a Taylor expansion; $\delta f(l):=f(x+l)-f(x)\simeq l f'(x)\propto l$ provided that $f'(x)$ exists and is finite.} Therefore, as we decrease $l$, the non-linear term $|{\bf R}_l|\sim l^2$ decreases \cite{EyinkAluie2006}. The spatial derivative, e.g., $\nabla\times{\bf R}_l$ which enters the renormalized induction equation (obtained by multiplying the bare induction equation by kernel $\phi({\bf r}/l)$ and integrating),
\begin{equation}\label{renormalizedinduction1}\partial_t {\bf B}_l=\eta\nabla^2{\bf B}_l+\nabla\times\Big({\bf v}_l\times{\bf B}_l-{\bf R}_l \Big),
  \end{equation}
  would scale as $|\nabla\times{\bf R}_l| \sim (l\times l)/l=l$ vanishing in the limit of small $l$. Everything is simple and physically expected; this is the situation in laminar flows.

Now suppose the Lipschitz continuity is lost. For instance, let us examine what would happen if $|{\delta \bf v}(l)|\propto l^{1/3}$ and $|{\delta \bf B}(l)|\propto l^{1/3}$. The derivatives e.g., of the form $|\nabla\times{\bf R}_l |\sim |{\bf R}_l|/l\sim (l^{1/3}\times l^{1/3})/l=l^{-1/3}$ will increase as we decrease the scale $l$! As we go to smaller scales, to make our measurements finer, unlike what happens in laminar flows, such derivatives in the equations of motion will make even larger contributions. Turbulence is a frequently encountered example in almost all astrophysical (and laboratory) plasmas where velocity and magnetic fields become non-Lipschitz. The empirical fact indicating the loss of Lipschitz continuity comes from the notion of {\it{anomalous dissipation}}.

\subsection{Anomalous Dissipation}
In a plasma with viscosity $\nu$, the kinetic energy is viscously dissipated at a rate (per mass) $\epsilon_{\nu}\equiv \nu|\nabla{\bf v}|^2$ while the magnetic energy is dissipated at a rate (per mass) $\epsilon_\eta\equiv \eta|\nabla{\bf B}|^2$. One may naively expect that these dissipation rates should vanish in the limit of vanishing viscosity and resistivity, i.e., $\lim_{\nu\rightarrow 0}\nu|\nabla{\bf v}|^2\rightarrow 0$ and $\lim_{\eta\rightarrow 0}\eta|\nabla{\bf B}|^2\rightarrow 0$. However, in a system of integral length $L$ and typical large scale velocity $V$, these limits correspond to large Reynolds number $Re:=LV/\nu$ and large magnetic Reynolds number $Re_m:=LV/\eta$. With large Reynolds numbers, the flow is extremely sensitive to perturbations and prone to become turbulent. Experiments and numerical simulations indicate that in turbulence, these energy dissipation rates remain finite even in the limit of vanishing viscosity and resistivity (see \cite{JV2019} and references therein): $\lim_{\nu\rightarrow 0}\nu|\nabla{\bf v}|^2\nrightarrow 0$ 
and $\lim_{\eta\rightarrow 0}\eta|\nabla{\bf B}|^2\nrightarrow 0$ (in turbulence).\footnote{Enhanced dissipation of kinetic energy is one of the characteristics of turbulent flows. G. I. Taylor was probably the first to suggest that kinetic energy can be dissipated even with infinitesimal viscosity. This idea of non-viscous turbulent dissipation in fully developed turbulence, called anomalous dissipation or the zeroth law of turbulence, was later explored by Kolmogorov and Onsager. For details and references, see \cite{DrivasEyink2017}.} Thus the spatial derivatives of velocity and magnetic fields should blow up to keep these rates finite; in other words, these {\it{dissipative anomalies}} imply divergent spatial derivatives for velocity and magnetic fields: $|\nabla {\bf v}|\rightarrow \infty$ and $|\nabla {\bf B}|\rightarrow \infty$. It is not hard to see that increments should scale as \citep{Eyink2018}
\begin{equation}
\label{anomalous3}
\delta v(l)\propto l^{1/3},
\end{equation}
and similarly
\begin{equation}
\label{anomalous4}
\delta B(l)\propto l^{1/3}.
\end{equation}
Note that with such singular spatial derivatives in turbulence, terms such as $\bf v.\nabla v$ and $\nabla\times (\bf v\times \bf B)$ which enter the Navier-Stokes and induction equations will also become ill-defined. In order to avoid such singularities, one can use the Lipschitz continuous renormalized fields, e.g., as defined by eq.(\ref{coarsegrain1}), at any (inertial) scale $l>0$. Equations of motion governing these renormalized fields, e.g., the renormalized induction equation given by eq.(\ref{renormalizedinduction1}), are thus "effective" field equations valid only on large (inertial range) scales $l$.

\section{Spontaneous Stochasticity}\label{SSS}
The scalings (\ref{anomalous3}) and (\ref{anomalous4}) have dramatic consequences for  determinism in classical physics. From eq.(\ref{anomalous3}), for example, it is obvious that
\begin{equation}\label{Kolmogorov}
{\delta v}(l)= dl(t)/dt\propto l^{1/3},
\end{equation}
which translates into the differential equation $dl(t)/dt=\alpha l^{1/3}$ with $\alpha=const.$ has a solution $l(t)=[l_0^{2/3}+{2\over 3}\alpha t ]^{3/2}$. Defining the characteristic time for the initial separation $l(0)=l_0$ as $t_0={3\over 2\alpha}l_0^{2/3}$, for long times, i.e., $t\gg t_0$, it follows that \begin{equation}\label{Richardson}
l^2(t)\sim {2\over 3}\alpha^3 t^3.
\end{equation}
As easily seen by setting $l_0=0$, which still results in $l^2(t)\sim t^3$ at long times, two particles starting from the {\it{same}} point separate to a finite distance! In other words, particle trajectories, i.e., the solution to the differential equation $dl(t)/dt= \delta v(l)$, are not unique. This is indeed a familiar example in the theory of differential equations for non-uniqueness: to guarantee a unique solution for 
\begin{equation}\label{SS1}
{d{\bf x}(t)\over dt}={\bf v(x},t), \;\;{\bf x}(0)={\bf x}_0,
\end{equation}the advecting velocity ${\bf v(x},t)$ should be Lipschitz continuous i.e., $|\delta{\bf v}(l)|\sim l^h$ with $h\geq 1$ (which basically means bounded spatial derivatives)\footnote{The real vector field ${\bf B}({\bf x})$ is H{\"o}lder continuous in ${{\bf x}}\in{\mathbb{R}}^n$ if $|{\bf B}(x) - {\bf B}(y) | \leq C| x - y|^h$ for some $C>0$ and $h>0$. If $h=1$, for any $x, y$, ${\bf B}$ is uniformly Lipschitz continuous. Also ${\bf B}$ is called H{\"o}lder singular if $0<h< 1$. A uniformly Lipschitz function $f(x)$, i.e., one which satisfies $|f(x)-f(y)|\leq C_f |x-y|^{h_f}$ for some $C_f>0$ with $h_f=1$, has a bounded derivative, i.e., $|f'(x)|< M$ for some $M>0$. In contrast, the derivative of a H{\"o}lder singular function $f$, i.e., one which satisfies $|f(x)-f(y)|\leq C_f |x-y|^{h_f}$ for some $C_f>0$ with $0<h_f<1$, can blow up; $|f'(x)|>\infty$.}.

Eq.(\ref{Kolmogorov}) is the well-known Kolmogorov \cite{Kolmogorov1941} scaling for turbulent velocity in the inertial range (i.e., range of scales where viscous effects are negligible). Eq.(\ref{Richardson}) is the famous Richardson's $2$-particle diffusion, as it was empirically discovered by Richardson in 1926 in his study of volcanic ash diffusion. The lack of uniqueness for solutions of eq.(\ref{SS1}) with Hölder continuous velocity ${\bf v(x},t)$ and the Richardson law are deeply connected to the concept of spontaneous stochasticity \citep{Bernardetal.1998}. An easy way to consider stochastic advection of fluid particles in a fluid with finite viscosity $\nu$ by a velocity ${\bf v}^\nu(\tilde{\bf x}(t), t)$ which is smooth in the dissipative scales:
\begin{equation}\label{Langevin1}
{d \tilde{\bf x}(t), t)\over dt} ={\bf v}^\nu(\tilde{\bf x}(t), t)+\sqrt{2D}\tilde {\boldsymbol\eta}(t),
\end{equation}
where $\tilde {\boldsymbol\eta}(t)$ is a Gaussian white noise with a strength $\sqrt{2D}$. The transition probability for a single fluid particle in a fixed non-random velocity realization $\bf v$ can be represented as the following path integral:
\begin{eqnarray}\nonumber
P_{\bf v}^{\nu, D}({\bf x}_f, t_f|&&{\bf x}_0, t_0)=\int_{{\bf x}(t_0)={\bf x}_0}{\mathcal D}{\bf x}\;\delta^3[{\bf x}_f-{\bf x}(t_f)]\\\nonumber
&&\times \exp\Big({-1\over 4D}\int_{t_0}^t d\tau |\dot{\bf x}(\tau)-{\bf v}^\nu({\bf x}(\tau),\tau) |^2 \Big).
\end{eqnarray}
The motivation to consider this problem comes from diffusion of a scalar field $\theta({\bf x},t)$, e.g., a concentration or temperature field, in the fluid with molecular diffusivity $D$, with advection-diffusion equation $\partial_t\theta+{\bf v}^\nu.\nabla\theta=D\nabla^2\theta$. The solution can be written using the Feynman-Kac formula
\begin{eqnarray}\nonumber
\theta({\bf x},t)&=&\int d^3x_0 \theta({\bf x}_0,t_0)P_{\bf v}^{\nu, D}({\bf x}_0,t_0|{\bf x}, t)\\\nonumber
&=&\int_{{\bf a}(t)={\bf x}} {\mathcal D}{\bf a}\;\theta({\bf a}(t_0),t_0)\\\nonumber
&&\;\times \exp\Big( {-1\over 4D}\int_{t_0}^t d\tau |\dot{\bf a}(\tau)-{\bf v}^\nu({\bf a}(\tau),\tau) |^2\Big),
\end{eqnarray}
 for $t_0<t$. In the limit of vanishing diffusivity, the noise term in eq.(\ref{Langevin1}) vanishes thus the physical expectation may be that in the limit $\nu, D\rightarrow 0$, the solution should become totally deterministic. In other words, the transition probability 
should collapse to a delta function, i.e., 
$${d{\bf x}\over dt}={\bf v(x}(t),t),\;\;\;\text{with}\;\;\;P_{\bf v}^{\nu, D}({\bf x}_f, t_f|{\bf x}_0,t_0)\rightarrow \delta^3[{\bf x}_f-{\bf x}(t_f)].$$
However, it has been shown \citep{Bernardetal.1998} that in the joint limit when viscosity and diffusivity both tend to zero, $\nu, D \rightarrow $, if the velocity field becomes non-smooth ${\bf v}^\nu\rightarrow {\bf v}$, the probability may tend to a non-trivial probability density thus Lagrangian trajectories may remain random! The breakdown in uniqueness of Lagrangian trajectories can appear in different physical limits for turbulent advection; See Sec. II in \citep{Eyink2011} and references therein. This phenomenon has been dubbed spontaneous stochasticity because of its similarity with spontaneous symmetry breaking in condensed matter physics and quantum field theory, where, for example, a ferromagnet may retain a non-vanishing magnetization even in the limit of vanishing external magnetic fields. Also, we note that spontaneous stochasticity differs from stochasticity commonly encountered in turbulence theory, associated to a random ensemble of velocity fields. Instead, the stochasticity in velocity field $\bf v$ in transition probability $P_{\bf v}^{\nu, D}$ is for a fixed (non-random) ensemble member $\bf v$ \cite{Eyink2011}. 

The theory of spontaneous stochasticity always starts with dynamics already stochastic, e.g. Landau-Lifschitz fluctuating hydrodynamics. The statement is that even in limit where the evolution equation formally becomes deterministic, the solutions remain random, with universal statistics that are independent of the underlying microscopic source of randomness. It is important to note that spontaneous stochasticity does not require an infinite Reynolds number, rather a large enough Reynolds number would suffice such that the limiting statistics are obtained. 

We will avoid the notion of magnetic field lines in the present work, but in passing we note that magnetic flux freezing has been traditionally formulated in terms of field lines \citep{Alfven1942, Newcomb1958}. Magnetic flux conservation in turbulent plasmas has been shown neither to hold in the conventional sense nor to be completely violated, instead it is valid in a statistical sense associated to the spontaneous stochasticity of Lagrangian particle trajectories \citep{Eyink2011, Eyinketal2013, Review2020}.

\subsection{Chaos and Super-Chaos}
Spontaneous stochasticity can also be called “super-chaos” which occurs due to formation 
of singularities and consequent divergence of Lyapunov exponents to infinity (see below). In such super-chaotic systems, e.g., turbulent flows, vanishingly small random perturbations can be propagated to much larger scales 
in a finite time. Note that this situation completely differs from simple chaos (the butterfly effect) where trajectories remain unique solutions of equations of motion and despite their explosive divergence, they never forget their initial separation. The fact that in Richardson diffusion particles starting from the {\it{same}} point separate to a finite separation at later times resembles the butterfly effect in chaos theory but in fact it is a completely different phenomenon. As discussed before, the continuity of velocity field means that 
\begin{equation}\label{Lip1}
|{\bf v}({\bf x}, t)-{\bf v}({\bf y}, t)|\leq H | {\bf x} -{\bf y}|^h
\end{equation}
for some real $H\geq 0$ and $0< h \leq 1$ (with $0<h<1$ implying Hölder singularity and with $h=1$ implying Lipschitz continuity). Consider the spatial separation of two arbitrary fluid particles ${\bf x}(t)$ and ${\bf y}(t)$ at time $t$, $\delta(t):=| {\bf x}(t) -{\bf y}(t)|$, which were initially separated by $\delta_0:=|{\bf x}(0)-{\bf y}(0)|$. Taking the time derivative of $\delta(t)$, we arrive at
$d\delta (t)/ dt\leq H[\delta(t)]^h$, with simple solution
\begin{equation}
\delta(t)\leq\Big[ \Delta_0^{1-h}+H(1-h)(t-t_0)  \Big]^{1\over 1-h}.
\end{equation}
In non-turbulent flows, the velocity field is Lipschitz, i.e., $h\rightarrow 1$, thus 
$\delta(t)\leq \delta_0 e^{ H(t-t_0)  }$. At long times, assuming a near equality, we get 

$$\delta(t)\simeq \delta_0e^{H(t-t_0)}.$$ 

Thus even with chaos, for smooth dynamical systems, there is at most exponential divergence of trajectories. The corresponding Lyapunov exponent
$$\lambda:=\lim_{t\rightarrow +\infty}\lim_{\delta_0\rightarrow 0} {1\over t-t_0}\ln\Big({\delta(t)\over\delta_0} \Big)$$
is the growth rate with $\lambda>0$ indicating chaos. hence, any small change in the initial
conditions (e.g., measurement errors) will be exponentially magnified. However, the initial conditions (the initial separations $\delta_0$) are never forgotten. In other words, similar to a laminar flow, in the limit when the initial separation vanishes, the final separation vanishes as well:
\begin{equation}\label{chaos}
\lim_{\delta_0\rightarrow 0} |{\bf x}(t)-{\bf y}(t) |\rightarrow 0, \;\;\text{(laminar/chaotic flow)}.
\end{equation}

On the other hand, a turbulent velocity field will be non-Lipschitz (or Hölder) with expoennt $0<h<1$. Hence in this case, we find $\Delta(t)\simeq \Big[H(1-h)(t-t_0) \Big]^{1\over 1-h}$, which implies that the information about initial conditions is completely lost! In other words, for long times, no matter how small the initial separation becomes, fluid particles separate super-linearly with time:
\begin{equation}\label{Richardson1}|{\bf x}(t)-{\bf y}(t) |\simeq t^{1\over 1-h},\;\;\text{(turbulent flow}),
\end{equation}
even in the limit $\delta_0\rightarrow 0$! This situation corresponds to an infinite Lyapunov exponent $\lambda=+\infty$, where the solutions remain unpredictable for all  times even with vanishing measurement errors in the initial conditions. For $h=1/3$, corresponding to the Kolmogorov scaling for velocity field \citep{Kolmogorov1941}, we recover the Richardson law \citep{Richardson1926} given by eq.(\ref{Richardson}).

\subsection{Alfv\'enic Wave Packets}

Similar to the argument following eq.(\ref{Kolmogorov}), we can use eq.(\ref{anomalous4}) to write ${\delta B}(l)=\delta V_A (l)= d{ l}(t)/dt\propto l^{1/3}$ or $d{ l}(t)/dt=\beta { l}^{1/3}$ (where assuming incompressibility, we have absorbed constant density to the definition of magnetic field). Thus, 
\begin{equation}
{ l}^2(t)\sim {2\over 3}\beta^3 t^3,
\end{equation}
which implies Richardson $2$-particle diffusion for magnetic disturbances moving with local Alfv\'en velocity. Trajectories of these "Alfv\'enic wave-packets" define magnetic path lines. Physically, we can imagine turbulence induces spatial fluctuations to magnetic field which can be regarded as small disturbances or "Alfv\'enic wave-packets" traveling with the Alfv\'en velocity ${\bf V}_A:={\bf B}/\sqrt{4\pi\rho}$ along the local field:
\begin{equation}\label{Wave-packets}
{d{\bf X}(t)\over dt}={\bf V}_A({\bf X},t), \;\;{\bf X}(0)={\bf X}_0.
\end{equation}

Hence, in turbulent flows, magnetic path lines diverge super-linearly with time. As we will see in \S\ref{sreconnection}, path lines and their divergence in time provide a natural way to study magnetic reconnection without invoking the notion of field lines whose motion in turbulence is a complicated process. Faraday's notion of magnetic field lines has played an extremely important role in electromagnetism and plasma physics. Yet, its utility becomes questionable in turbulent flows where they cannot be defined as smooth curves evolving continuously in time as unique solutions of a differential equation. Their advection and diffusion through a turbulent plasma might deceptively look simple but in fact constitute a complicated problem in magnetohydrodynamics. Path lines provide an alternative tool with a simpler dynamics.

Assuming an incompressible flow with constant density $\rho$, we will absorb the density into the definition of magnetic field and write ${d{\bf X}(t)\over dt}={\bf B}({\bf X},t)$ as the equation defining path lines. It is easy to see that similar to velocity field, the anomalous dissipation of magnetic field, thus its  Hölder singularity as indicated by eq.(\ref{anomalous4}), will lead to Richardson diffusion of Alfv\'enic wave-packets, which can be called "magnetic quasi-particles" in the language of modern field theories. Consider two wave-packets separated by $|\bf {X}(t)-{\bf Y}(t)|$ at time $t$, initially separated by $\Delta_0:=\Delta(0)$. Using the continuity condition for magnetic field ${\bf B(X}(t), t)$ similar to (\ref{Lip1}), 
\begin{equation}\label{Lip2}
|{\bf B}({\bf X}, t)-{\bf B}({\bf Y}, t)|\leq G | {\bf X} -{\bf Y}|^g,
\end{equation}
with $G\geq 0$ and $0< g\leq 1$ and repeating the argument which led to eqs. (\ref{chaos}) and (\ref{Richardson1}), we obtain $\Delta(t)=\Delta_0 e^{G t}$ implying 
$$\lim_{\Delta_0\rightarrow 0} |{\bf X}(t)-{\bf Y}(t) |\rightarrow 0, \;\;\text{(laminar/chaotic flow)}$$
with finite Lyapunov exponents $\lambda=const.$
In turbulence, however, we have 
$$\lim_{\Delta_0\rightarrow 0}|{\bf X}(t)-{\bf Y}(t) |\simeq t^{1\over 1-g}\simeq t^{3/2},\;\;\text{(turbulent flow})$$
with infinite Lyapunov exponents $\lambda\rightarrow +\infty.$

An argument analogous to what we advanced for the stochastic scalar diffusion can be applied to magnetic field diffusion. The solution of the advection-diffusion equation for the magnetic field, i.e., the induction equation\footnote{The fluctuation-dissipation theorem in statistical physics dictates the presence of a fluctuating term for any dissipative term such as $\eta\nabla^2 {\bf B}$ in equations of motion. For example, this can be achieved by adding a random term $\nabla.{\boldsymbol \tau}_\eta({\bf x},t)$ where the tensor ${\boldsymbol \tau}_\eta$ is a fluctuating field, e.g., prescribed as a Gaussian random noise, whose strength depends on resistivity and temperature. Despite the importance of the random term on small dissipative scales, it can be ignored in the inertial range of turbulence which we are interested in the present work. Analogously, and more famoulsy, such a fluctuating term corresponding to viscous dissipation should also be added to the Navier-Stokes equation; see e.g., \cite{Bandaketal2022}. The resulting equation is called fluctuating Navier-Stokes equation in fluctuating hydrodynamics \cite{Zarate2006}.}, 
\begin{equation}
\partial_t{\bf B}+{\bf v.\nabla B=B.\nabla v}+\eta\nabla^2{\bf B},
\end{equation}
with initial condition ${\bf B}(t_0)={\bf B}_0$ can be written as a path integral
\begin{eqnarray}\nonumber
{\bf B(x},t)&=&\int_{{\bf a}(t)={\bf x}} {\mathcal D}{\bf a}\; {\bf B}_0[{\bf a}(t_0)].{\mathcal J}({\bf a},t)\\\nonumber
&&\times \exp\Big( {-1\over 4\eta}\int_{t_0}^t d\tau |\dot {\bf a}(\tau)-{\bf v}^\nu({\bf a}(\tau),\tau)  |^2 \Big)
\end{eqnarray}
where ${\mathcal J}$ is a $3\times 3$ matrix and $\bf B$ is interpreted as a row-vector \citep{Eyink2011}. The condition ${\bf a}(t)=\bf x$ corresponds to solutions of
\begin{equation}\label{Langevin2}
{d \tilde{\bf a}(\tau)\over d\tau} ={\bf v}(\tilde{\bf a}(\tau), \tau)+\sqrt{2\eta}\;\tilde {\boldsymbol\eta}(t),\;\;\tilde{\bf a}(t)=\bf x,
\end{equation}
integrated backward in time from $\tau=t$ to $\tau=t_0$. 
Similar to the discussion we had for the scalar field diffusion, one may naively apply Laplace  asymptotic method to the path integral in the limit of vanishing resistivity and viscosity and expect to recover a single deterministic solution in the limit of $\nu, \eta\rightarrow 0$. However, in this limit, the solution may remain random if the velocity and magnetic fields become non-smooth, which is what encountered in turbulence. This situation is intimately related to the breakdown of the traditional magnetic flux freezing; for a more detailed discussion see \citep{Eyink2011}.

\section{Magnetic Reconnection}\label{sreconnection}
Reconnection can be defined as a rapid change, i.e., on time scales much faster than resistive, in magnetic field pattern which accelerates and heats up the plasma \citep{Review2020}.
Physically, one expects that such rapid changes in magnetic field direction in a region of plasma should correspond to explosive divergence of Alfv\'enic wave-packets. One way of looking at this phenomenon is to consider the 
Alfv\'enic wave-packets. As we will see presently, this approach suggests that the failure of Sweet-Parker model \citep{Sweet1958, Parker1957} in astrophysics is due to our assumption that magnetic diffusion is a "normal" diffusion process due to resistivity whereas in real turbulent flows, turbulence causes a super-diffusion (Richardson diffusion) at much larger scales which affects reconnection. 

In a fluid of diffusivity $D$ (solvent), the mean square separation of diffusing particles (solute) grows as $\delta_D^2(t)\simeq D t$ with time\footnote{Linearity (super-linearity) in time implies normal (super) diffusion. Resistive diffusion is a Taylor diffusion which concerns the rms displacement of a Lagrangian particle from its initial position as opposed to Richardson's diffusion where the rms separation of two particles is considered.} Similarly, resistivity $\eta$ causes diffusion of Alfv\'enic wave-packets whose mean square separation grows as $\delta_\eta^2(t)\simeq \eta t$. One can understand the classical Sweet-Parker reconnection very easily in terms of the diffusion of these wave-packets. In a current sheet of length $\Delta$ and thickness $\delta$, where reconnection proceeds with a typical speed $V_R$, the conservation of mass leads to $\Delta \times V_R=\delta \times V_A$ with the Alfv\'en velocity $V_A$. Magnetic diffusion on the scale of current sheet's thickness $\delta$ implies 
$\delta^2\simeq \eta t$ 
where $t\simeq \Delta/V_A$ is the time  Alfv\'enic wave-packets require to traverse the current sheet's length $\Delta$. Thus, combining these two basic results, we recover the famous Sweet-Parker reconnection speed
\begin{equation}\label{SP1}V_R\simeq \sqrt{\eta{V_A\over \Delta}}.
\end{equation}
This speed is of course too slow compared with observations \citep{Review2020}. The reason is that almost all astrophysical flows are turbulent; much faster turbulent diffusion at much larger (inertial) scales, instead of resistive diffusion at much smaller scales, dominates magnetic field evolution and reconnection.  

Turbulence is a complicated phenomenon, however, turbulent reconnection is much simpler to understand than many (unrealistic) non-turbulent models which are based on instabilities for example. As discussed before, in hydrodynamic turbulence, the mean square separation $\delta_\nu^2(t)$ of pairs of Lagrangian fluid particles grows super-linearly with time i.e., $\delta_\nu^2(t)\simeq \epsilon_\nu t^3$
(Richardson diffusion). In magnetohydrodynamic turbulence, the mean square separation of Alfv\'enic wave-packets grows as 
\begin{equation}\label{Richardson10}
\delta_R^2(t)\simeq \epsilon_\eta t^3,
\end{equation}
corresponding to the Richardson diffusion of wave-packets. Combining $\epsilon_\eta=u_L^4/V_A L_i$, where $L_i$ is the energy injection scale and $u_L$ is the (isotropic) injection velocity, with $\delta^2\simeq\epsilon_\eta t^{3}$, one obtains the Lazarian-Vishniac reconnection speed
\begin{equation}\label{LV99}
V_R\simeq V_A\sqrt{L/L_i}M_A^2,
\end{equation}
with Mach number $M_A$.\footnote{This is Eq.(3.12) in \citep{Lazarianetal2015} which was obtained using an argument based on "field line diffusion".} 

In passing, we also note that the Richardson diffusion of wave-packets can be also formulated in terms of magnetic field lines: field line wandering in MHD turbulence constantly undergo reconnection on all inertial scales. This new type of dispersion of field lines is sometimes called the Richardson diffusion in space, as opposed to the original Richardson dispersion in time \cite{ELV2011}. Also, note that Richardson diffusion of wave-packets, or magnetic field lines studied in e.g., \cite{ELV2011}, is intimately related to the violation of traditional magnetic flux freezing \cite{Alfven1942} in turbulent flows. Our emphasis is on a picture independent of magnetic field lines thus we will not consider the problem of flux freezing in terms of field lines here; for detailed discussions and numerical evidence see e.g., \citep{Eyink2011, ELV2011, Eyinketal2013}.

Almost all non-turbulent reconnection models proposed to enhance the Sweet-Parker rate rely on small-scale plasma effects\footnote{The Petschek model \cite{Petschek1964} relies on an $X$-shape geometry to reduce the effective length of the current sheet to enhance reconnection rate.}, e.g., represented by a non-ideal term $\bf P$ in the generalized Ohm's law, ${\bf E+v\times B=P}$, which leads to the induction equation
\begin{equation}\label{renormalizedinduction2}\partial_t {\bf B}_l=\nabla\times\Big({\bf v}_l\times{\bf B}_l-{\bf R}_l-{\bf P}_l \Big),
\end{equation}
where we have also included the resistive electric field in $\bf P$. Reconnection is driven by the non-ideal terms $(\nabla\times{\bf R}_l)/B_l$ and $(\nabla\times{\bf P}_l)/B_l$ which act as source terms for the differential equations governing magnetic field direction $\hat{\bf B}_l:={\bf B}_l/B_l$ \citep{JV2019, Review2020}. However, $|\nabla\times{\bf R}_l|$ grows for $l$ decreasing in the turbulence inertial range, until it becomes comparable to the plasma non-ideal term $|\nabla\times{\bf P}_l|$ at the turbulence microscale $l_d$. For $l\gg l_d$, 
$|\nabla\times{\bf R}_l|\gg |\nabla\times{\bf P}_l|$ (see e.g., \citep{Eyink2015} and references therein). Hence, in the presence of turbulence, plasma effects essentially play no role in reconnection. This fact, indicated by the early work of Lazarian and Vishniac \cite{LV99} more than two decades ago, has been confirmed both by simulations and observations, yet the reconnection community seems to have lagged behind by appealing to instabilities such as tearing modes. While plasma non-ideal effects affect reconnection near electron- and ion-scales, these effects are completely  irrelevant to reconnection  at scales larger than the ion gyroradius.

\section{Final Remarks and Conclusions}

The main message of this short review is that magnetic fields especially in astrophysics are usually turbulent, study of which requires concepts such as regularization/renormalization, stochastic dynamics and effective field theories. Utilizing such tools from other areas such as quantum field theories and condensed matter physics, which may seem exotic in plasma physics, does not necessarily lead to a more complicated approach. In fact, in some problems like reconnection, it turns out that simple notions such as super-linear Richardson diffusion \cite{ELV2011,Eyink2018,JVV2019} or magnetic disturbances taken as (quasi) particles in the context of dynamical system theory \cite{Jafari2024} can provide a more intuitive picture.

Turbulence is a super-chaotic state plagued with indeterminism and singularities, which is totally different from chaos that is inherently deterministic. This is an important matter because the common approach to reconnection in plasma physics literature is based on either plasma instabilities and non-idealities, e.g., \cite{ZweibelYamada2009, Yamadaetal2010}, or chaos theory, e.g., \cite{Boozer2018,Boozer2023} and references therein. However, observations and simulations all point to the ubiquity of turbulence which can be driven by plasma instabilities; magnetic reconnection and/or external sources e.g., stellar winds and  jets in astrophysics (for observational evidence, see e.g., Sec II A in \cite{Review2020} and references therein). In particular we note that the overall effect of non-ideal plasma mechanisms, e.g., Hall effect, tearing modes instabilities etc., on large scale astrophysical reconnection is negligible, somehow resembling negligible quantum effects in classical systems.

Magnetohydrodynamic turbulence and reconnection seem to be interconnected in the sense that one may derive and intensify the other. Turbulent reconnection is not only fast and independent of plasma effects, it should in fact be regarded as an inseparable part of MHD turbulence as it constantly occurs on all inertial scales in a random manner. The original formulation of turbulent reconnection was based on stochastic field line wandering \cite{LV99}, later refined mathematically using the idea of Richardson diffusion of field lines \citep{ELV2011, Eyinketal2013, Eyink2015, Lazarianetal2019, Review2020}. However, as discussed in \S\ref{sreconnection}, turbulent reconnection can be easily formulated and understood in terms of Alfv\'enic wave-packets as (quasi) particles moving along magnetic path lines. The latter approach has two advantages: first, it avoids the complication of field lines motion through a turbulent plasma as these curves are not persistent dynamical entities in time and their diffusion has been source of confusion and misunderstanding in the reconnection literature. The second advantage is the possibility of easily formulating magnetic topology and its rate of change in the context of a dynamical system theory. One implication is that if reconnection is defined as magnetic topology-change, it can be fast only in turbulent flows where both reconnection and topology-change are driven by spontaneous stochasticity, independent of any plasma effects \cite{Jafari2024}.

\bibliography{main}{}
\end{document}